\documentclass[aps,prl,twocolumn,superscriptaddress]{revtex4}
\usepackage{graphicx }
\usepackage{bm}
\usepackage{amsmath}

\begin{document}
\title{Experimental Evidence for Giant Vortex States in a Mesoscopic Superconducting Disk}
\author{A. Kanda}
\email[]{kanda@lt.px.tsukuba.ac.jp} \affiliation{Institute of
Physics and Tsukuba Research Center for Interdisciplinary Material
Science (TIMS), University of Tsukuba, Tsukuba 305-8571, Japan}
\author{B. J. Baelus}
\email[]{ben.baelus@ua.ac.be} \affiliation{Institute of Materials
Science, University of Tsukuba, Tsukuba 305-8573, Japan}
\affiliation{Departement Fysica, Universiteit Antwerpen (Campus
Drie Eiken), Universiteitsplein 1, B-2610 Antwerpen, Belgium}
\author{F. M. Peeters}
\affiliation{Departement Fysica, Universiteit Antwerpen (Campus
Drie Eiken), Universiteitsplein 1, B-2610 Antwerpen, Belgium}
\author{K. Kadowaki}
\affiliation{Institute of Materials Science, University of
Tsukuba, Tsukuba 305-8573, Japan}
\author{Y. Ootuka}
\affiliation{Institute of Physics and Tsukuba Research Center for
Interdisciplinary Material Science (TIMS), University of Tsukuba,
Tsukuba 305-8571, Japan}
\date{\today}

\begin{abstract}
Response of a mesoscopic superconducting disk to perpendicular
magnetic fields is studied by using the
multiple-small-tunnel-junction method, in which transport
properties of several small tunnel junctions attached to the disk
are measured simultaneously. This allows us for the first
experimental distinction between the giant vortex states and
multivortex states. Moreover, we experimentally find magnetic
field induced rearrangement and combination of vortices. The
experimental results are well reproduced in numerical results
based on the nonlinear Ginzburg-Landau theory.
\end{abstract}


\maketitle

The appearance of vortices in various quantum systems, such as
superconductors, superfluids and Bose-Einstein condensates, is an
intriguing phenomenon in nature. A conventional quantum vortex is
singly quantized, having a core where the value of the order
parameter decreases to zero, while its phase changes by $2\pi$
when encircling the core. Recently, an important breakthrough was
established by the observation of doubly quantized vortex lines in
superfluid $^3$He-A\cite{Blaauwgeers00}. For superconductors
expectations are even more spectacular. In macroscopic type-II
superconductors a triangular lattice of single flux quanta is
formed, whereas two kinds of fundamentally new vortex states have
theoretically been predicted in mesoscopic superconductors where
the sample size approaches the size of
Cooper-pairs\cite{schweigert8,palacios8,palacios0,baelus1}; (i)
multivortex states (MVSs) with a unique spatial arrangement of
singly quantized vortices, and (ii) multiply quantized or giant
vortex states (GVSs) with a single core in the
center\cite{mosgvs,bru9}.

Although several experimental techniques have been developed for
observing these novel
states\cite{bru9,geim7,geim8,geim0,chibotaru0,chibotaru1,hata3},
none of them has been able to make a clear distinction between
MVSs and GVSs. In this Letter, we present the first experimental
evidence for the existence of GVSs and MVSs in a circular disk,
and demonstrate magnetic-field induced MVS-GVS and MVS-MVS
transitions. Our results are in good agreement with the
theoretical prediction based on the nonlinear Ginzburg-Landau
(G-L) theory.

Here we used the multiple-small-tunnel-junction (MSTJ) method, in
which several small tunnel junctions with high tunnel resistance
are attached to a mesoscopic superconductor to simultaneously
detect small changes in the local density of states (LDOS) under
the junctions\cite{kanda2,kanda4}. Since the LDOS depends on the
local supercurrent density, the MSTJ method gives us information
on the distribution of the supercurrent, which reflects the
detailed vortex structure inside the disk.

\begin{figure}
\includegraphics{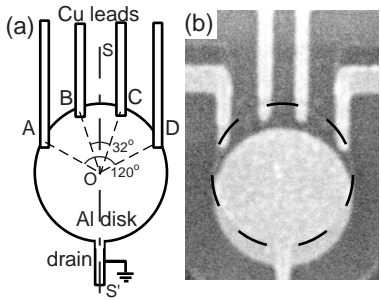}
\caption{\label{fig1}Schematic view (a) and scanning electron
micrograph (b) of our sample. Four normal-metal (Cu) leads are
connected to the periphery of a superconducting Al disk (diameter
$=1.5$ $\mu$m, thickness $= 33$ nm) through small tunnel junctions
with area $\approx 0.01$ $(\mu$m$)^2$. The disk is directly
connected to an Al drain lead. This structure was fabricated using
e-beam lithography followed by double-angle evaporation of Al and
Cu. After the Al film was deposited, the surface of the Al film
was slightly oxidized to provide the tunnel barrier. Most of the
Al disk, indicated by the dashed circle, is covered with a Cu film
(bright regions). We expect that the Cu film will not have any
serious influence on the superconductivity of the Al disk because
of the insulating AlO$_x$ layer between them. }
\end{figure}

Figure \ref{fig1} shows a schematic drawing and a scanning
electron microscopy (SEM) image of the sample. Four normal-metal
(Cu) leads are connected to the periphery of the superconducting
Al disk through highly resistive small tunnel junctions, A, B, C,
and D. The sample is designed to be symmetrical with respect to
the central axis $SS'$. The angles $\angle AOD$ and $\angle BOC$
are 120 and 32 degrees, respectively. Although junctions A and D
and junctions B and C ideally have the same area and tunnel
resistance, small differences actually exist between them. The
normal-state tunnel resistance at 8 K was 40 and 33 k$\Omega$ for
junctions A and D, and 17 and 25 k$\Omega$ for junctions B and C,
respectively. The radius of the disk $R$ was 0.75 $\mu$m and the
disk thickness $d$ was 33 nm. The disk was directly connected to
an Al drain lead. To prevent oxidation of the disk in the air, we
covered the Al surface with Ge (thickness: 28 nm), which becomes
insulating at low temperatures \cite{Ge-comment}. All the
above-mentioned processes were performed in a single vacuum with a
base pressure of $2 \times 10^{-8}$ Pa. The superconducting
coherence length $\xi$ was estimated to be 0.15 to 0.19 $\mu$m
from the residual resistance of the Al films prepared in the same
way. The superconducting transition temperature was 1.3 - 1.4 K.

In the measurement, we fixed the current flowing through each
junction to a small value, typically 100 pA, and measured
simultaneously the voltages between each of the four Cu leads and
the drain lead, while sweeping the perpendicular magnetic field at
a typical rate of 20 mT/min. Here, the current $I$ is related to
the voltage $V$ through the superconducting LDOS $N_s$:
\begin{equation}
I=\frac{1}{eR}\int\nolimits_{0}^{eV}\frac{N_s(E)}{N_n}dE \text{
(for }T\rightarrow 0),
\end{equation}
where $N_n$ is the normal density of states. Especially, at $T =
0$, $B=0$ and $I \rightarrow 0$, $V=\Delta /e$. Variations in the
LDOS are related to variations in the superconducting density
$\left| \Psi \right|^2$.

\begin{figure}
\includegraphics[width=0.7\linewidth] {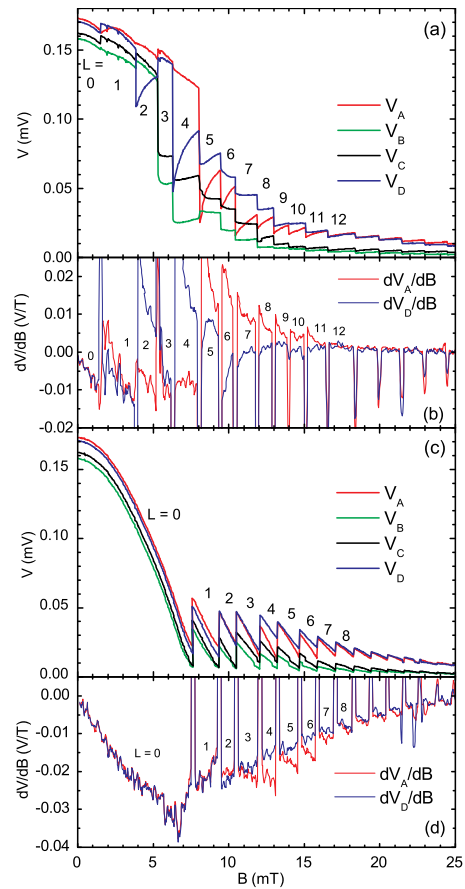}
\caption{\label{fig2} (a) Variation of voltages at junctions A, B,
C, and D in a decreasing magnetic field. The current through each
junction is 100 pA. Temperature is 0.03 K. (b) Differential
voltage $dV/dB$ for junction pairs at symmetrical positions, A and
D. (c)(d) The same as (a) and (b), respectively, but for
increasing magnetic fields.}
\end{figure}

Figures \ref{fig2}(a) and (c) show the change of the voltages at
$I = 100$ pA in decreasing and increasing magnetic fields,
respectively. $V_A$, $V_B$, $V_C$, and $V_D$ denote the voltages
at junctions A, B, C, and D, respectively. The magnetic-field
dependence of the voltage originates from (i) smearing of the
energy gap due to pair-breaking by the magnetic field, and (ii) a
decrease of the energy gap because of the supercurrent
\cite{tinkham}.

The former leads to a moderate monotonic decrease in voltage as
the strength of the magnetic field increases, so the rapid change
in voltage comes from the latter. Especially, each voltage jump
corresponds to a transition between different vortex states with a
vorticity change of $\pm 1$
\cite{schweigert8,palacios8,palacios0}. This allows us to identify
the vorticity $L$ (the number of the flux quanta in the sample) as
shown in each figure \cite{note1}. Note that the difference either
between $V_A$ and $V_D$ or $V_B$ and $V_C$ at $B = 0$ mainly comes
from a slight asymmetry in the junction resistance, which would
also affect the characteristics in all magnetic-field ranges. To
make voltage comparison easier, $dV/dB$ is also displayed in Figs.
\ref{fig2}(b) and (d).

Here, we focus on the features of the voltages in the symmetric
junctions, $V_A$ and $V_D$ \cite{note3}. In decreasing magnetic
field (Fig. 2(b)), remarkable differences are found in $dV_A/dB$
and $dV_D/dB$ for $L = 2$ and 4 to 11. This difference between
$dV_A/dB$ and $dV_D/dB$ indicates that the supercurrent below
junction A is essentially different from the one below junction D,
which excludes an axially symmetric vortex distribution of the GVS
and is characteristic of the MVS. This allows for an unparalleled
determination of the magnetic field for which the vortex state is
an MVS. For increasing magnetic fields (Fig. \ref{fig2}(d)), the
difference in $dV/dB$ is relatively large between $L = 4$ and 6,
which is also due to the MVS formation.

\begin{figure}
\includegraphics[width=0.7\linewidth]{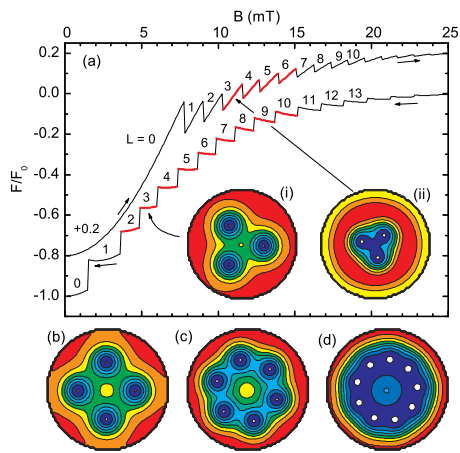}
\caption{\label{fig4} (a) Calculated free energy $F$ for a disk
with $R=5.0\xi$, $d = 0.1\xi$, and $\kappa = 0.28$, normalized by
the $B=0$ value $F_0$, for decreasing and increasing magnetic
fields. For the sake of clarity the free energy for increasing
field is shifted over $+0.2F_{0}$. Red and black segments indicate
MVS and GVS, respectively. The insets show the contour plots of
the Cooper-pair density for the $L = 3$ state (i) at $B = 6.0$ mT
and (ii) at $B = 11.0$ mT, corresponding to decreasing and
increasing magnetic field, respectively. Red (blue) regions
correspond to high (low) values of the Cooper-pair density. The
Cooper-pair density for decreasing magnetic field is also shown
for (b) the $L=4$ state at $B=7.2$ mT, (c) the $L=6$ state at
$B=9.3$ mT, and (d) the $L=9$ state at  $B=13.0$ mT. }
\end{figure}

This simple distinction between GVSs and MVSs is supported by a
numerical simulation. Figure \ref{fig4} shows the free energy for
a disk with $R = 5.0\xi$, $d = 0.1\xi$, and the G-L parameter
$\kappa = 0.28$ as calculated within the framework of the
nonlinear G-L theory. This theoretical analysis is based on a
fully self-consistent numerical solution of the coupled G-L
equations, taking into account demagnetization effects. A more
detailed description of the theoretical model can be found in Ref.
\cite{schweigert8}. The thickness was adjusted to obtain the best
agreement with the experimentally obtained transition field
between $L = 0$ and 1 states \cite{kanda4,note2}. Theoretically,
MVSs nucleate for vorticity $L = 2$ to 10 for decreasing magnetic
fields and $L = 3$ to 6 for increasing magnetic fields. Thus, the
theoretical calculations confirm the identification of the GVS and
MVS by the MSTJ method except for $L = 3$ ($L = 11$), where
theoretically the state is predicted to be an MVS (GVS) while
experimentally a GVS (MVS) was inferred.

The disagreement for $L =3$ originates from the junction
configuration. The contour plots in Fig. \ref{fig4} show examples
of the theoretically expected vortex configuration for the MVSs.
The $L = 3$ state (insets of Fig. \ref{fig4}(a)) has trigonal
symmetry, which agrees with the angle $\angle AOD = 120$ degree.
Thus, the voltage difference for the $L = 3$ MVS is significantly
decreased for the A-D junction pair, concealing the $L=3$ MVS.
Although this kind of symmetry induced effect is also expected to
appear in $L=6$ and $L=9$ MVSs (Figs. \ref{fig4}(c) and (d)), the
$dV/dB$ difference is significant for $L = 6$ and for $L=9$, as
shown in Figs. \ref{fig2}(b) and (d). The difference between the
experiment and theory could be attributed to the effects of
defects, i.e., the stabilization of a different vortex
configuration or distortions of the vortex configurations caused
by defects. Note that in such mesoscopic disks different vortex
distributions with the same total vorticity $L$ are possible
\cite{baelus4-2}. For example, a state with $L$ vortices on one
shell or a state with $L-1$ vortices on a shell and one in the
center may become (meta-)stable. The free energy difference
between such states with the same vorticity is very small and even
the smallest defect (inside the disk or at the boundary of the
disk) can influence the vortex distribution. The experimentally
observed MVS for $L=11$ could also be attributed to the effect of
defects.

Actually, the influence of defects was noticeable for particular
vorticities; e.g., for the $L=0$ state in both Figs. \ref{fig2}(a)
and (c), all curves are parallel to each other, showing that a
uniform supercurrent is flowing along the disk periphery. This
means that there is no crucial defect {\it near the junctions}. On
the other hand, for the $L=1$ state, where only one vortex exists
in the disk, curves are not exactly parallel. This indicates that
the vortex is not exactly at the center of the disk, presumably
because of a defect close to (but not at) {\it the disk center}.
This will be confirmed in the discussion below. The small
differences in vortex state transition fields between experiment
and theory can also be attributed to the effect of defects
\cite{braverman9}.

\begin{figure}
\includegraphics[width=0.85\linewidth]{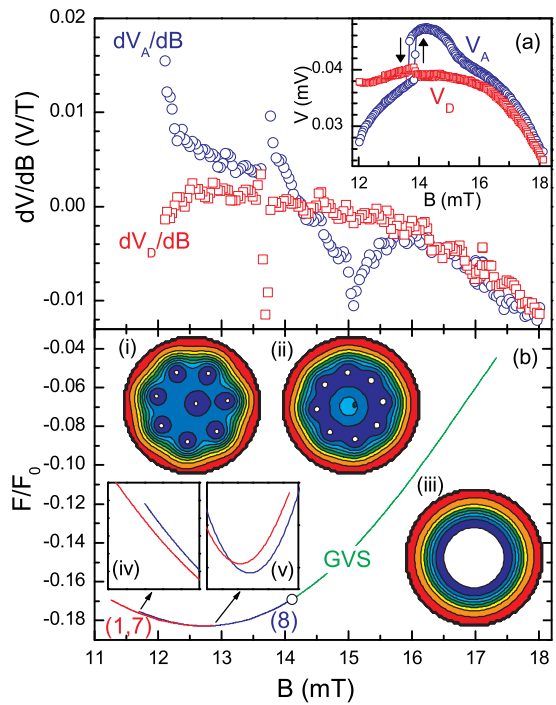}
\caption{\label{fig5} (a) The main panel shows the differential
voltages $dV_A/dB$ and $dV_D/dB$ for the $L = 8$ state in
decreasing magnetic field. The inset shows the measured variation
of voltages $V_A$ and $V_D$ for the $L = 8$ state. The arrows show
the direction of the magnetic-field sweep. Small hysteresis is
seen at the voltage jump (around 13.7 mT). (b) Comparison of the
calculated free energy of the different (meta-) stable states with
$L = 8$ as a function of the applied magnetic field; i.e, the
(1,7)-state (red curve), the (8)-state (blue curve) and the GVS
(green curve). To simulate a defect near the center, a circular
hole with radius $0.1\xi$ is inserted at a distance $0.2\xi$ from
the disk center. The insets (i)-(iii) present the Cooper-pair
density of the (1,7) and the (8)-state at $B = 12.5$ mT, and the
GVS at $B = 14.6$ mT. Insets (iv) and (v) show the transitions
between the (1,7)-state and the (8)-state in more detail. }
\end{figure}

In increasing magnetic fields, the $dV/dB$ difference for MVS
formation (Fig. \ref{fig2}(d)) is relatively small. This results
from the position of the vortices. The vortices in increasing
magnetic fields (e.g., inset (ii) of Fig. \ref{fig4}(a)) are
situated more to the center in comparison with those for
decreasing magnetic fields (e.g., inset (i) of Fig.
\ref{fig4}(a)), leading to less variation of the supercurrent near
the disk periphery. Similarly, we also attribute the smaller
differences in $dV/dB$ in MVSs with larger $L$ (Fig.
\ref{fig2}(b)) to less variation of the supercurrent along the
disk periphery in comparison with smaller $L$, as shown in Figs.
\ref{fig4}(b,c,d).

For $L=2$ and $L= 7$ to 11, the type of the vortex state is
different in decreasing and increasing magnetic fields. This
implies the existence of a MVS-GVS transition at these
vorticities, which has been predicted theoretically
\cite{schweigert8}, but has never been observed experimentally.
Figure \ref{fig5}(a) shows the entire $L=8$ state, obtained by
changing the sweep direction of magnetic field. The difference
between $dV_A/dB$ and $dV_D/dB$ is remarkable below 15.8 mT,
indicating that the state is an MVS, while at larger fields
$dV_A/dB$ and $dV_D/dB$ coincide, indicating a GVS. Note that the
observed MVS-GVS transition is a continuous one and is not
accompanied by hysteresis, correspond to the theoretical
prediction \cite{schweigert8,palacios0}. Moreover, small voltage
jumps with hysteresis observed around 13.7 mT [see the inset of
Fig. \ref{fig5}(a)] indicate a transition between two vortex
configurations with a different arrangement of the 8 vortices,
which might be related to the presence of a defect near the sample
center. For comparison, we calculated the different vortex
configurations with $L = 8$ in a disk with a weak defect near the
center (Fig. \ref{fig5}(b)). At low fields, we find two stable
MVSs with $L = 8$; one is the (1,7)-state, in which 1 vortex
exists near the center pinned by the defect and 7 vortices are
located on a shell [inset (i)], and the other is the (8)-state, in
which all 8 vortices are arranged on a shell [inset (ii)]. At
higher fields the GVS with $L = 8$ is found to be most stable
[inset (iii)]. With decreasing field the GVS transits into the
(8)-state at $B = 14.1$ mT and then into the (1,7)-state at $B =
11.7$ mT. With increasing field the (1,7)-state transits into the
(8)-state at $B = 12.9$ mT, and then into the giant vortex state
at $B = 14.1$ mT. Note that the MVS-GVS transition at $B = 14.1$
mT is a continuous one (second order transition), in agreement
with the experimental observation, while the transition between
the two MVSs is discontinuous and hysteretic [see insets (iv) and
(v) of Fig. \ref{fig5}(b)], as was also the case in the experiment
[see the inset of Fig. \ref{fig5}(a)]. Also note that the
(1,7)-state is theoretically stable only in the presence of weak
defects, which is consistent with the observation in the $L = 1$
state, as discussed above.

Thus, the major experimental results for this particular disk are
successfully explained by the calculation taking into account a
single strong defect near the disk center. But as seen in the
figures, small discrepancies exist in fields for transitions
between different vortex states obtained in experiment and theory.
Also, different disks with the same geometry have a little bit
different transition fields \cite{kanda4}. These might come from a
distribution of much weaker defects. Additional experiments on a
number of similar disks will be required to resolve this matter.
However, this is beyond the scope of the present letter.

In conclusion, we have studied the magnetic response of a
mesoscopic superconducting disk by using the MSTJ method. By
comparing the voltages at symmetrical positions, we experimentally
determine the type of vortex states: GVS or MVS. We also observed
the MVS-MVS and MVS-GVS transitions with a fixed vorticity. The
results agree with theoretical predictions based on the G-L
theory.

Finally, we want to remark that our method can be applied to other
geometries such as squares and triangles, in which antivortices
might be stabilized for some vorticities \cite{chibotaru0,
chibotaru1, bekaert4,morelle4}.

\begin{acknowledgments}
This work was partially supported by the University of Tsukuba
Nanoscience Special Project, the 21st Century COE Program of MEXT,
the JSPS Core-to-Core Program, the Flemish Science Foundation
(FWO-Vl) and the Belgian Science Policy. B. J. B. acknowledges
support from JSPS and FWO-Vl.
\end{acknowledgments}


\begin{thebibliography}{99}
\bibitem{Ge-comment} The effect of the Ge layer on the Al surface oxidization has not been confirmed experimentally.
\bibitem{note1} Small voltage jumps in $V_B$ at 0.010 mV and in $V_C$ at 0.013 mV seen in Figs. \ref{fig2}(a) and (c) do not correspond to vortex state transitions, but are due to errors in the voltage measurement at these voltages, presumably because of change in the measurement range in our digital multimeters.
\bibitem{note3} In $V_B$ and $V_C$, the effects discussed in the text were less pronounced, presumably because of the short separation between the two leads.
\end{thebibliography}
\end{document}